\documentclass[longauth]{aa}  
\pdfoutput=1
\usepackage{graphicx}
\usepackage{txfonts}
\usepackage{natbib}
\usepackage{rotating}

\bibpunct{(}{)}{;}{a}{}{,}
\usepackage{hyperref}
\usepackage{xcolor, colortbl}
\begin{document} 


\title{Young [$\alpha$/Fe]-enhanced stars discovered by CoRoT and APOGEE:
What is their origin?}

\author{C. Chiappini\inst{1,2},
F. Anders\inst{1,2}, T. S. Rodrigues\inst{2,3,4}, A. Miglio\inst{5}, J. 
Montalb\'an\inst{4}, B. Mosser\inst{6}, 
L. Girardi\inst{2,3}, M. Valentini\inst{1}, A. Noels\inst{7}, T. 
Morel\inst{7}, I. Minchev\inst{1}, M. Steinmetz\inst{1}, 
B. X. Santiago\inst{2,8},  M. Schultheis\inst{9}, M. 
Martig\inst{10}, L. N. da Costa\inst{2,11},  M. A. G. Maia\inst{2,11}, 
 C. Allende Prieto\inst{12,13}, R. de Assis Peralta\inst{6}, S. 
Hekker\inst{14,15}, N. Theme\ss{}l\inst{14,15}, T. Kallinger\inst{16}, R. A. Garc\'\i a\inst{17}, S. Mathur\inst{18}, F. Baudin\inst{19}, 
T. C. Beers\inst{20}, K. Cunha\inst{11}, P. Harding\inst{21}, J. Holtzman\inst{22}, S. Majewski\inst{23}, Sz.~M{\'e}sz{\'a}ros\inst{24,25},  
D. Nidever\inst{26}, K. Pan\inst{22,27}, R. P. Schiavon\inst{28},
M. D. Shetrone\inst{29}, D. P. Schneider\inst{30,31}, K. Stassun\inst{32}
}

\authorrunning{Chiappini, Anders et al.}    
\titlerunning{Young [$\alpha$/Fe]-enhanced stars discovered in a CoRoT-APOGEE 
sample}

\institute{
Leibniz-Institut f\"ur Astrophysik Potsdam (AIP), An der Sternwarte 16, 
14482 Potsdam, Germany
\and{Laborat\'orio Interinstitucional de e-Astronomia, - LIneA, Rua Gal. Jos\'e 
Cristino 77, Rio de Janeiro, RJ - 20921-400, Brazil}
\and{Osservatorio Astronomico di Padova -- INAF, Vicolo dell'Osservatorio 5, 
I-35122 Padova, Italy}
\and{Dipartimento di Fisica e Astronomia, Universit\`a di Padova, Vicolo 
dell'Osservatorio 3, I-35122 Padova, Italy}
\and{School of Physics and Astronomy, University of Birmingham, Edgbaston, 
Birmingham, B15 2TT, United Kingdom}
\and{LESIA, Universit\'{e} Pierre et Marie Curie, Universit\'{e} Denis Diderot, 
Obs. de Paris, 92195 Meudon Cedex, France}
\and{Institut d'Astrophysique et de Geophysique, All\'ee du 6 ao\^ut, 17 - 
B\^at. B5c, B-4000 Li\`ege 1 (Sart-Tilman), Belgium}
\and{Instituto de F\'\i sica, Universidade Federal do Rio Grande do  Sul, Caixa 
Postal 15051, Porto Alegre, RS - 91501-970, Brazil}
\and{Observatoire de la Cote d'Azur, Laboratoire Lagrange, CNRS UMR 7923, B.P. 
4229, 06304 Nice Cedex, France}
\and{Max-Planck-Institut f\"ur Astronomie, K\"onigstuhl 17, D-69117 
Heidelberg, Germany}
\and{Observat\'orio Nacional, Rua Gal. Jos\'e Cristino 77, Rio de Janeiro, RJ 
- 20921-400, Brazil}
\and{Instituto de Astrof\'{\i}sica de Canarias, 38205 La Laguna, Tenerife, Spain}
\and{Universidad de La Laguna, Departamento de Astrof\'{\i}sica, 38206 La Laguna, Tenerife, Spain}
\and{Max-Planck-Institut f\"ur Sonnensystemforschung, Justus-von-Liebig-Weg 3, 
37077 G\"ottingen, Germany}
\and{Stellar Astrophysics Centre, Department of Physics and Astronomy, Aarhus 
University, Ny Munkegade 120, DK-8000 Aarhus C, Denmark}
\and{Institut f\"ur Astronomie, Universit\"at Wien, T\"urkenschanzstr. 17, 
Wien, Austria}
\and{Laboratoire AIM, CEA/DSM -- CNRS - Univ. Paris Diderot -- IRFU/SAp, Centre de Saclay, 91191 Gif-sur-Yvette Cedex, France}
\and{Space Science Institute, 4750 Walnut Street Suite  205, Boulder CO 80301, USA}
\and{Institut d\'{}Astrophysique Spatiale, UMR8617, CNRS, Universit\'{e} Paris XI, B\^{a}timent 121, 91405 Orsay Cedex, France}
\and{Dept. of Physics and JINA-CEE: Joint Institute for Nuclear Astrophysics -- Center for the Evolution of the Elements, Univ. of Notre Dame, Notre Dame, IN 46530 USA} 
\and{57 Department of Astronomy, Case Western Reserve University, Cleveland, OH 44106, USA}
\and{New Mexico State University, Las Cruces, NM 88003, USA}
\and{Department of Astronomy, University of Virginia, PO Box 400325, Charlottesville VA 22904-4325, USA}
\and{ELTE Gothard Astrophysical Observatory, H-9704 Szombathely, Szent Imre herceg st. 112, Hungary}
\and{Department of Astronomy, Indiana University, Bloomington, IN 47405, USA}
\and{Dept. of Astronomy, University of Michigan, Ann Arbor, MI, 48104, USA}
\and{Apache Point Observatory PO Box 59, Sunspot, NM 88349, USA}
\and{Astrophysics Research Institute, Liverpool John Moores University, IC2, Liverpool Science Park 146 Brownlow Hill Liverpool L3 5RF, UK}
\and{Mcdonald Observatory, University of Texas at Austin, HC75 Box 1337-MCD, Fort Davis, TX 79734, USA}
\and{Department of Astronomy and Astrophysics, The Pennsylvania State University, University Park, PA 16802}
\and{Institute for Gravitation and the Cosmos, The Pennsylvania State University, University Park, PA 16802}
\and{Vanderbilt University, Dept. of Physics \& Astronomy, VU Station B 1807, Nashville, TN 37235, USA}
}

\date{Received \today; accepted ...}

\abstract{
We report the discovery of a group of apparently young CoRoT red-giant stars 
exhibiting enhanced [$\alpha$/Fe] abundance ratios (as determined from APOGEE spectra) with respect 
to solar values. Their existence is not explained by standard chemical evolution models of 
the Milky Way, and shows that the chemical-enrichment history of the Galactic 
disc is more complex. 
We find similar stars in previously published samples for which isochrone-ages 
could be reliably obtained, although in smaller relative numbers.
This might explain why these stars have not previously  received attention. 
The young [$\alpha$/Fe]-rich stars are much more numerous 
in the CoRoT-APOGEE (CoRoGEE) inner-field sample than in any other high-resolution sample 
available at present because only CoRoGEE can explore the inner-disc regions and 
provide ages for its field stars. 
The kinematic properties of the young [$\alpha$/Fe]-rich stars are not clearly thick-disc like,
despite their rather large distances from the Galactic mid-plane. 
Our tentative interpretation of these and previous intriguing observations in the Milky Way is that these stars  
were formed close to the end of the Galactic bar, near corotation -- a 
region where gas can be kept inert for longer times than in other regions 
that are  more frequently shocked by the passage of spiral arms. 
Moreover, this is where the mass return from older
inner-disc stellar generations is expected to be highest (according to an 
inside-out disc-formation scenario), which additionally dilutes the in-situ gas.
Other possibilities to explain these observations (e.g., a recent gas-accretion event) are also discussed.
}


\maketitle

\section{Introduction} 

One of the pillars of Galactic Archaeology is the use of stellar [$\alpha$/Fe] 
abundance ratios as an indirect age estimator: [$\alpha$/Fe]-enhancement is an 
indication that a star has formed from gas enriched by core-collapse 
supernovae; longer-timescale polluters, such as supernovae of type Ia 
or asymptotic giant-branch stars, did not have sufficient time to enrich the interstellar 
medium \citep{Pagel2009, Matteucci2001}. High-resolution spectroscopy of the 
solar neighbourhood stars, for which Hipparcos parallaxes are available \cite[e.g.][]{Haywood2013}, have indeed 
shown this paradigm to work well. One of the best examples is the very local 
($d<25$ 
pc) volume-complete sample of solar-like stars by \citet[and references therein]{Fuhrmann2011}, for which it was possible to obtain robust isochrone ages for a 
small number of subgiants, which confirmed that stars exhibiting [$\alpha$/Fe]-enhancements 
were all older than $\sim$10 Gyr and identified them as thick-disc stars. Fuhrmann's 
data also show a clear chemical discontinuity in the [$\alpha$/Fe] vs. [Fe/H] 
plane, which can be interpreted as the result of a star-formation gap between the thick and thin discs \citep*{Chiappini1997, Fuhrmann2011}. 

However, as we demonstrate in this Letter, $\alpha$-enhancement is no guarantee 
that a star is actually old. Only recently has it become possible to obtain 
more precise ages for field stars far beyond the solar circle, thanks to 
asteroseismology, with CoRoT \citep{Baglin2006} and {\it Kepler} 
\citep{Gilliland2010}. Even more important, the CoRoT mission allows for age 
and distance determination of stars spanning a wide range of Galactocentric 
distances, as shown by \cite{Miglio2013a, Miglio2013b}. The latter authors have 
shown that when asteroseismic scaling relations are combined 
with photometric information, mass and age can be obtained to a precision 
of about 10\% and 30\%, respectively, even for distant objects\footnote{The quoted 
uncertainties in \cite{Miglio2013a} were computed assuming 
global seismic parameter uncertainties from \cite{Mosser2010}. Similar age 
uncertainties are found here, despite using spectroscopic information -- as we have 
now adopted not only individual uncertainties but also a more conservative 
uncertainty estimate for the seismic parameters (details can be found in Anders 
et al. 2015, in prep.).}. High-resolution spectroscopy 
of the seismic targets plays a key role, not only allowing for more precise ages 
and distances, but also providing full chemical and kinematical information. 

We have initiated a collaboration between CoRoT and APOGEE (the Apache Point Observatory Galactic Evolution Experiment; Majewski et al., in prep.).  
APOGEE is a high-resolution ($\mathrm{R \sim}$22,000) infrared survey ($\lambda=1.51-1.69\,\mu$m) and part of the Sloan Digital Sky Survey III \cite[SDSS-III]{Eisenstein2011}, which uses the Sloan 2.5 m telescope 
\citep{Gunn2006}. Here, we analyse data from the SDSS-III Data Release 12 (DR12; 
\citealt{Alam2015}), which contains 690 red-giant stars in the CoRoT fields 
LRa01 and LRc01 from an ancillary APOGEE campaign.  

The CoRoT-APOGEE sample (CoRoGEE) studied here is briefly described in Sect. 
\ref{obs}, while a more detailed description can be found in Anders et al. (2015, in 
prep.; hereafter A15). 
The latter paper describes the analysis performed to extract the main stellar properties for this sample, such as masses, radii, ages, distances, extinctions, and kinematic parameters. 
 The authors also present some immediate results that can 
be obtained with the CoRoGEE sample, such as the variation of the disc 
metallicity gradient with time or age-chemistry relations outside the 
solar vicinity.
In the present Letter, we focus on a group of stars which, 
despite being enhanced in [$\alpha$/Fe], appear to be relatively young. 
Because these stars, at first sight, challenge the currently accepted paradigm, 
we carry out several tests to consolidate our assigned ages and abundances in our companion paper.
In Sect.~\ref{results} we identify the young high-[$\alpha$/Fe] stars and
describe their main properties, and in Sect.~\ref{origin} we discuss possible interpretations for their origin.
Our main conclusions are summarised in Sect.~\ref{conclusions}.

\section{Observations}\label{obs}

The CoRoT data we employed are a subset of the larger sample 
analysed by \citet{Miglio2013a}. Red-giant oscillation spectra have been 
analysed as in \cite{Mosser2010}. The global seismic parameters $\Delta\nu$ and 
$\nu_{\mathrm{max}}$ were measured following the method described in 
\cite{Mosser2009}. When possible, a more precise determination of the large spacing was derived from the correlation of the power spectrum with the universal red-giant oscillation pattern \citep{Mosser2011}. Outliers to the 
$\Delta\nu$-$\nu_{\mathrm{max}}$ relation, which would correspond to unrealistic 
stellar masses, were excluded.

These targets were observed by APOGEE, and 
the high-resolution infrared spectra were analysed with the APOGEE Stellar 
Parameter and Chemical Abundances Pipeline (ASPCAP; \citealt{Meszaros2013}, Garc\'\i a P\'erez, al., in prep.). Here, we 
adopted internally calibrated DR12 abundances
(\citealt{Holtzman2015}; see more details in A15.). 

We used the Bayesian code PARAM 
\citep{daSilva2006} to estimate stellar parameters. Masses, ages, 
distances, and extinctions were
obtained with an updated version of the code \citep{Rodrigues2014},
which uses the combined photometric, seismic, and spectroscopic
information to compute the probability density functions of these stellar properties. The final sample adopted here contains 622
red giant stars from the CoRoT LRa01 ($(l,b)=(212,-2)$) and LRc01 ($(l,b)=(37,-7)$) fields, for which a) high-quality spectroscopic criteria 
are fulfilled (APOGEE spectra with SNR $>$ 90, 4000 K $< \mathrm{T_{eff}} < $5300 K, 1 $< \log g < $3.5), b) the PARAM code converged, and c) the seismic and calibrated spectroscopic  $\log g$ are consistent within 0.5 dex.
For this sample, statistical uncertainties of about 0.02 dex in $\log g$, 3\% in radius, 8\% in mass, 25\% in age, and 2.5\% in distance were obtained (more details can be found in 
A15.). As a caveat, stellar ages might still be affected by systematic uncertainties related to 
different stellar models and helium content, among other sources of errors (\citealt{Lebreton2014, Lebreton2014a, Martig2015}; Miglio et al., in prep.). 

Our dataset is complemented with similar information coming from two
other high-resolution samples for which isochrone ages were available, 
the F \& G solar-vicinity stars of \citet*{Bensby2014}, and the Gaia-ESO first 
internal data release of UVES spectra analysed in \citet{Bergemann2014}.
The total number of stars in each each sample is reported in Table~\ref{sampletable}.

\section{Discovery of young [$\alpha$/Fe]-rich stars in the Galactic disc}
\label{results}

\begin{figure*}
\centering
\includegraphics[width=.92\textwidth]{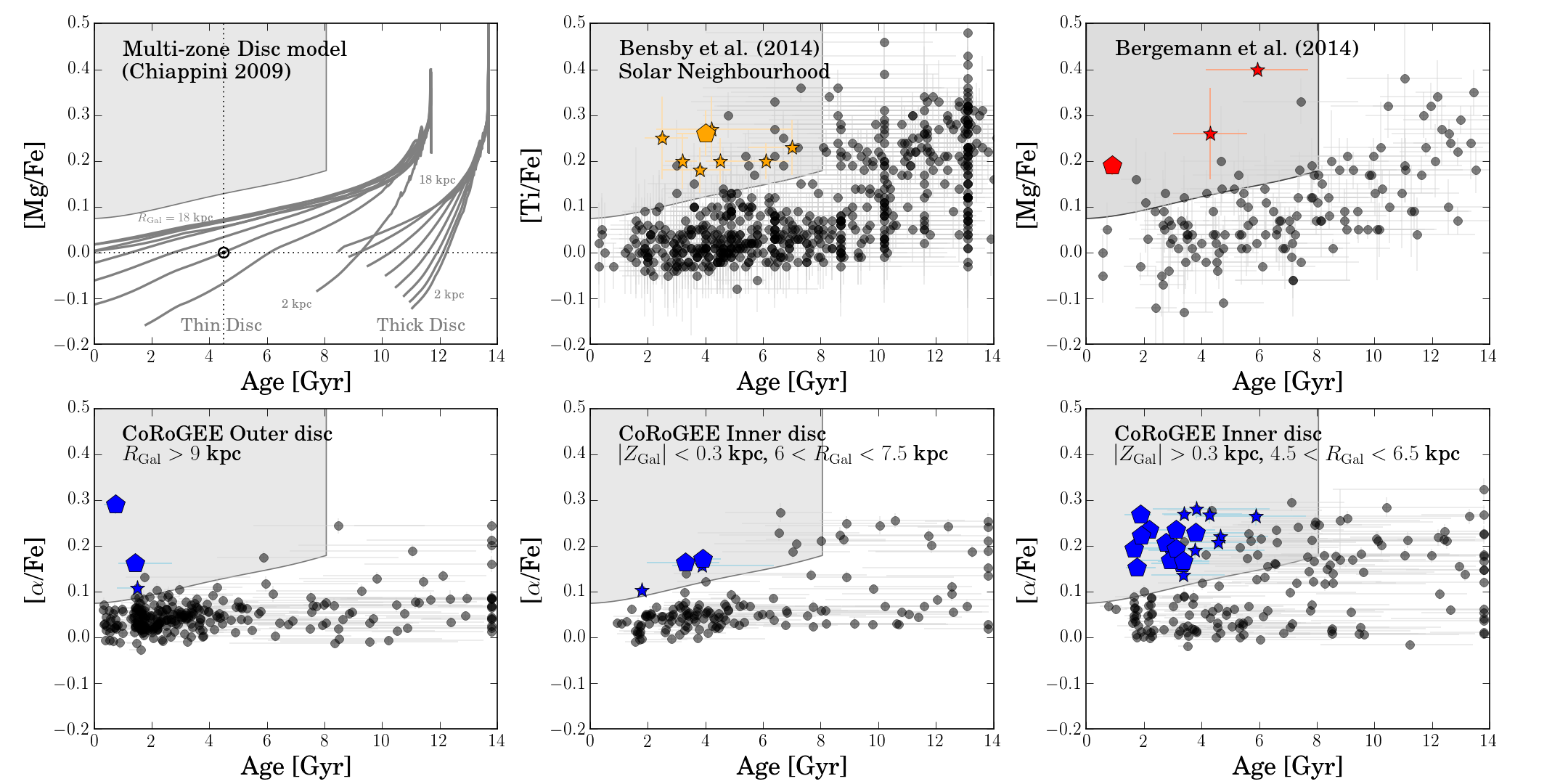}
\caption{Age-[$\alpha$/Fe] relation in different regions of the Galactic 
disc. {\it Upper left panel}: The grey curves indicate the predictions of the 
multi-zone Galactic chemical-evolution model of \citet{Chiappini2009} for the thin and thick discs, where different tracks were calculated for different 
Galactocentric annuli situated between 2 and 18 kpc from the Galactic Centre. 
The solar position is indicated in the diagram for the 6~kpc curve, the distance of the most 
probable birth position of the Sun \citep*{Minchev2013}.
Within these models, it is not possible to explain stars that 
fall into the grey-shaded region of the diagram: young, [$\alpha$/Fe]-enhanced 
stars. The grey shadings provide a heuristic estimate of the typical 
$N\sigma$ ($N=1,2,3$) uncertainties in [Mg/Fe] and age.
{\it Upper middle and right panels}: The solar cylinder data from \citet[][middle panel,]{Bensby2014} and the Gaia-ESO survey (\citealt{Bergemann2014}; right panel) 
show a clear correlation between isochrone-derived age estimates and relative 
[$\alpha$/Fe] abundances. Stars whose age and abundance estimates are 1$\sigma$-incompatible with any of 
the chemical evolution curves are represented by stars; 
2$\sigma$-outliers are represented by pentagons.
{\it Lower panels}: The same diagram for the CoRoT-APOGEE sample.
{\it Left}: the LRa01 outer-disc field. {\it Middle}: the LRc01 inner-disc field, 
close to the Galactic plane ($|Z_{\mathrm{Gal}}|<0.3$ 
kpc, $R_{\mathrm{Gal}}>6.0$ kpc). {\it Right}: the LRc01 field, below the Galactic 
plane ($Z_{\mathrm{Gal}}<-0.3$ kpc, 
$R_{\mathrm{Gal}}<6.5$ kpc). In this region, the 
fraction of young $\alpha$-enhanced stars is much larger than in all other 
regions. Considering {\it normal stars} alone, the 
age-[$\alpha$/Fe] relation is much flatter than locally because the CoRoT stars span a 
wide range in Galactocentric distances.}
\label{Fig1}
\end{figure*}

\begin{figure}\centering
 \includegraphics[width=.44\textwidth]{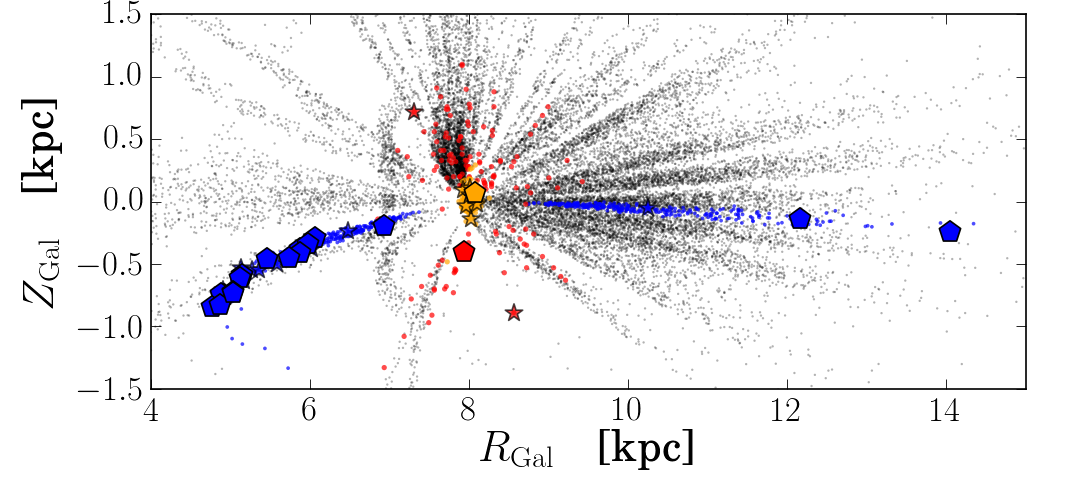}
  \caption{Location of the APOGEE high-quality sample of \citet{Anders2014} in 
a $Z_{\mathrm{Gal}}$ vs. $R_{\mathrm{Gal}}$ plane (grey points). Also shown are 
the locations of the CoRoGEE stars ({\it blue}), the subgiant stars from \citet[][{\it red}]{Bergemann2014}, and the \citet{Bensby2014} solar-vicinity dwarf stars ({\it orange}). As in Fig. \ref{Fig1}, the discovered young [$\alpha$/Fe]-rich stars are represented by the pentagons and stars.}
\label{Fig2}
 \end{figure}

Figure \ref{Fig1} presents the age-[$\alpha$/Fe] abundance relation for two local high-resolution spectroscopy samples: GES-UVES \citep{Bergemann2014} and 
\citet*{Bensby2014}. The lower row shows the same relation for our CoRoGEE sample split into {\it1)} outer-field (LRa01)
stars, {\it 2)} inner-field (LRc01) stars with $Z_{\mathrm{Gal}}<0.3$ kpc, and {\it 3)} inner-field stars with $Z_{\mathrm{Gal}}>0.3$ kpc. The 
latter is necessary because for the inner field, stars of different heights below 
the mid-plane span different Galactocentric distance ranges. This behaviour is a 
consequence of the way the LRc01 CoRoT field was positioned (see Fig~\ref{Fig2}; for more information 
on the population content of the LRc01 and LRa01 fields, see 
\citealt{Miglio2013a}). 

We also show in Fig. \ref{Fig1} (upper-left panel) the predictions
for the [Mg/Fe] vs. age chemical evolution of \citet{Chiappini2009} for different Galactocentric 
annuli of the thick and thin discs. These models assume that the thick disc was 
formed on much shorter timescales and with a higher star formation efficiency 
than the thin disc.
The shaded area corresponds to a parameter space not covered by a standard 
chemical evolution model of the thick and thin discs. Figure \ref{Fig1} demonstrates that while 
most of the data can be explained by standard chemical evolution models plus
 observational uncertainties (most probably accompanied by significant 
radial mixing, as discussed in \citealt{Chiappini2009} and 
\citealt*{Minchev2013, Minchev2014}), several stars are found to possess rather 
high [$\alpha$/Fe] ratios, despite their young ages, and hence cannot be 
accounted for by the models. These stars are depicted as stars (1$\sigma$-outliers) and pentagons (2$\sigma$-outliers) in all 
figures. The young [$\alpha$/Fe]-rich stars are more numerous in the inner field (see Fig.~\ref{Fig2} and Table~\ref{sampletable}).

Table~\ref{sampletable} shows the occurrance rates of 
young [$\alpha$/Fe]-rich stars in 
the different analysed samples. Interestingly, there is a sudden rise in the 
fraction of young [$\alpha$/Fe]-rich stars when smaller Galactocentric distances are sampled (which is the case of the CoRoT LRc01 field for $Z_{\mathrm{Gal}}>0.3$ 
kpc), and the absence of these stars in the \citet{Fuhrmann2011} 
sample, as well as other less volumed-confined samples such as \citet{Ramirez2007} -- which might be due to a statistical effect.

\begin{table}
\caption{Abundance of young $\alpha$-enhanced stars (y$\alpha$r) in recent 
high-resolution spectroscopic surveys.}
{\footnotesize
\setlength{\tabcolsep}{5pt}
\begin{tabular}{lcrc}\hline\hline
  \multicolumn{1}{c}{Sample} &
  \multicolumn{1}{c}{$R_{\mathrm{Gal}}$$^a$} &
  \multicolumn{1}{c}{N$^b$} &
  \multicolumn{1}{c}{1-$\sigma$ / 2-$\sigma$ y$\alpha$r} \\
 & [kpc] &   & \\
 \hline
Fuhrmann$^c$, $d<25$ pc & 8 & 424 & 0 / 0 \\
Bensby et al.$^d$ & 8 & 714 & 8 (1.1\%) / 1 (0.1\%)\\
GES$^e$, $|Z_{\mathrm{Gal}}|<0.3$ kpc & 6~--~9 & 55 & 0 / 0 \\
GES$^e$, $|Z_{\mathrm{Gal}}|>0.3$ kpc & 6~--~9 & 91 & 3 (3.3\%) / 1 (1.1\%)\\
LRa01$^f$ & 9~--~14 & 288 & 3 (1.0\%) / 2 (0.7\%)\\
LRc01$^f$, $|Z_{\mathrm{Gal}}|<0.3$ kpc  & 6~--~7.5 & 151 & 4 (2.6\%) / 2 (1.3\%)\\
LRc01$^f$, $|Z_{\mathrm{Gal}}|>0.3$ kpc & 4~--~6.5 & 183 & 21 (11.5\%) / 13 (7.1\%)\\
APOKASC$^g$ & 7~--~8 & 1639 & 14 (0.8\%) \\
\hline\end{tabular}\\
}
\tablefoot{
(a) Galactocentric range covered by different samples (b) N = 
total number 
of stars in the sample, (c) The volume-complete sample of 
\citet{Fuhrmann2011}, (d) Hipparcos volume \citep*{Bensby2014}, (e)  {\it i}DR1 
\citep{Bergemann2014}, (f) CoRoGEE, this work -- see Appendix for detailed information on each star, (g) \citet{Martig2015}. Outliers 
were defined in a different manner than in the present work.
}
\label{sampletable}
\end{table}

\begin{figure}\centering
 \includegraphics[width=.45\textwidth]{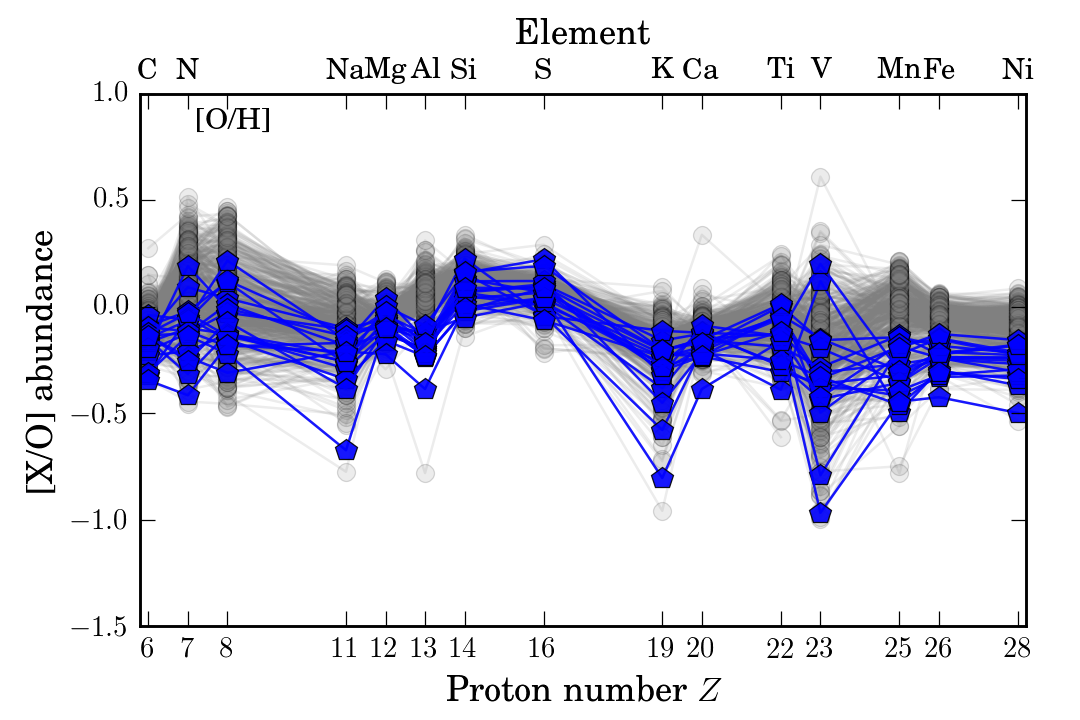}
 \label{parallel_chem}
 \caption{Chemical-abundance patterns relative to oxygen for the CoRoGEE 
stars marked as chemically peculiar in Fig.~\ref{Fig1} (blue hexagons, 2$\sigma$-outliers in the age-[$\alpha$/Fe] diagram). The chemical 
abundance pattern of the rest of the CoRoGEE sample is presented in grey for comparison. 
}
\end{figure}

The young [$\alpha$/Fe]-rich stars cover a wide range of stellar parameters ($4200\,\mathrm{K}\,< T_{\mathrm{eff}}<5100\,\mathrm{K}, 1.7<\log g<2.7$; see also Fig. 10 of \citealt{Martig2015}). The abundance pattern of these stars compared to the entire 
CoRoGEE sample is displayed in Fig.~\ref{parallel_chem}. These stars are compatible 
with being formed from a gas that has not been processed by many stellar generations, as 
indicated by the systematically lower abundance of iron-peak elements (lower 
contribution of type Ia supernovae to the chemical enrichment), as well as by 
the lower [N/O] and [C/O] abundance ratios (further indicating a mild contribution 
from intermediate-mass stars) with respect to the bulk of the CoRoGEE sample. 
However, when we restrict the comparison to stars with [O/H] $< -0.2$, no significant differences are detected any more.

We also investigated the kinematic properties of the young [$\alpha$/Fe]-rich stars. Despite their [$\alpha$/Fe] enhancements, many of them exhibit thin-disc 
like kinematics (although biased to hotter orbits because the inner CoRoT 
field samples Galactocentric distances below $\sim$5 kpc only at larger 
distances from the mid-plane, $Z_{\mathrm{Gal}}>0.3$ kpc). As
a result of sample 
selection effects, stars with small Galactocentric distances are only reachable 
at large distances from the mid-plane and should not be mistaken for 
genuine thick-disc stars.

\begin{figure}\centering
 \includegraphics[width=0.45\textwidth]{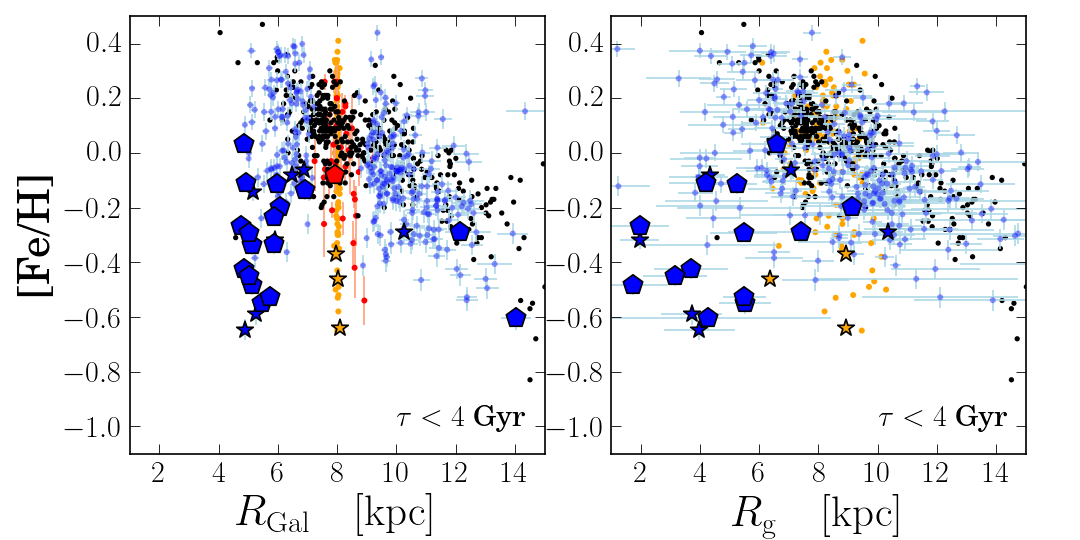}
 \caption{Radial [Fe/H] distribution ({\it left}: as a function of  Galactocentric distance $R_{\mathrm{Gal}}$, {\it right}: w.r.t. the guiding radius $R_g$)
over the extent of the Galactic disc (4-14 kpc range).
As in Fig. \ref{Fig2}, the CoRoGEE sample is shown in blue, the \citet{Bergemann2014} stars in red, and the \citet{Bensby2014} sample in orange. Again, hexagons and stars represent the young [$\alpha$/Fe]-enhanced stars defined in Fig.~\ref{Fig1}. The locations of Galactic cepheids ({\it black}; 
data from \citealt{Genovali2014}) are also indicated. }
\label{gradients}
\end{figure}

Focusing on the youngest stars (ages younger than 4 Gyr), where most of the 2$-\sigma$ 
outliers are found (see Fig.~\ref{Fig1}), we checked the locus of the young 
[$\alpha$/Fe]-rich stars in the [Fe/H] vs. Galactocentric distance diagram 
(Fig.~\ref{gradients}, left panel) and in the [Fe/H] vs. guiding radius 
diagram (Fig.~\ref{gradients}, right panel). Similar to 
\cite*{Minchev2014}, 
we estimated the guiding-centre radius of a stellar orbit using the 
approximation ${{R_g}} = \frac{L_z}{v_c} = \frac{v_{\phi} \cdot R_{\mathrm{Gal}} 
}{v_c}$, with $L_z$ being the angular momentum, $v_{\phi}$ the $\phi$-component 
of the space velocity in a Galactocentric cylindrical coordinate frame, and 
$v_c$ the circular velocity at the star position -- which for 
simplicity we assumed to be constant and equal to 220 $\mathrm{km~s}^{-1}$ (see A15 for more 
details). 

It is clear that most of the anomalous stars tend to be metal
poor and to have small guiding radii ($R_g \lesssim$ 6~kpc - dashed line in Fig.~\ref{gradients}). This is 
also the case of the young [$\alpha$/Fe]-rich stars in the other two more local 
samples.  In particular, a large number of these anomalous objects appear 
near the corotation region (with the caveat that there are large uncertainties in the 
estimate of the guiding radii). It is expected that as the age increases, more 
of these stars can also be found farther away from the corotation radius 
because radial migration would have had enough time to displace
them from their 
birth position \citep*{Minchev2014}. A larger age-baseline is discussed 
in A15, where we focus on the time evolution of abundance gradients. 

\section{What is their origin?}\label{origin}

One possible interpretation is that the young $[\alpha$/Fe]-rich stars might be evolved blue stragglers, that is, binary mergers.
These have a higher mass and thus look like a young population. However, these stars should be present in all directions, at all metallicities, but in smaller numbers (see discussion in \citealt{Martig2015}).

The young $\alpha$-rich stars appear to have been born from a relatively pristine 
gas, with metallicities above [Fe/H] $\sim -$0.7 (see Fig.~\ref{gradients}, and 
Table~\ref{A1}).
One possibility is that these are objects formed from a recent gas accretion 
event. One caveat here is that outliers are also 
present in older age bins, suggesting that the processes responsible 
for creating these stars have been continuously working during the Milky Way 
evolution. A more plausible interpretation is to assume that the region near the bar corotation is the site for the formation of the young [$\alpha$/Fe]-rich 
stars. In this region, gas can be kept inert for longer times
than in other 
regions that are more often shocked by the passage of the spiral arms
\citep{Bissantz2003,Combes2014}. Additional dilution is expected from gas restored from the death of old low-mass stars in this inner-disc region \citep*{Minchev2013}. 

If this interpretation holds and the process is still taking place 
in a region near the end of the Galactic bar, we also expect to find 
young metal-poor, [$\alpha$/Fe]-enhanced stars in that same region of the Galactic plane. Interestingly, there are some intriguing young objects in the MW that might be related to the same phenomenon: 
{\it a)} the puzzling low-metallicity supergiants located near 
the end of the Galactic bar (\citealt{Davies2009, Davies2009a}, see discussion 
in \citealt{Genovali2014} and \citealt{Origlia2013}),  
{\it b)} the young [$\alpha$/Fe]-enchanced stars reported by \cite{Cunha2007} near the Galactic Centre, and, 
{\it c)}  the unusual Cepheid BC Aql which, 
despite being young (Whitelock, priv. comm.) and located at R$_{\mathrm{Gal}} \sim$5 kpc, is also [$\alpha$/Fe]-enhanced and metal-poor \citep{Luck2011a}. 
Other Cepheids, recently discovered 
far from the Galactic plane on the 
opposite side of the Galaxy \citep{Feast2014}, 
also appear to be young 
(i.e., their period-age relations are compatible with ages $\lesssim$ 130 Myr).
 
Within our framework, we expect similar stars to have been forming in that same 
region (i.e., near the bar corotation) for the past 4-5 Gyr. As extensively 
discussed by \cite*{Minchev2013, Minchev2014}, stars born at 
the corotation radius have a high probability of being expelled to an outer 
region via radial migration. This result suggests that the mechanism proposed 
here could have a strong effect on the thin disc by contaminating the 
entire 
disc with this metal-poor and [$\alpha$/Fe]-rich population and that it might be 
related to the observed [Fe/H]$\sim-$0.7 \emph{floor} in the abundance 
gradients. One possible observable signature of this process might be the 
intermediate-age $\alpha$-enhanced open clusters found by \citet*[][and 
references therein]{Yong2012}.

\section{Conclusions}\label{conclusions}

In this Letter we reported the discovery of young [$\alpha$/Fe]-enhanced stars in 
a sample of CoRoT stars observed by APOGEE (CoRoGEE). These stars have a 
lower iron-peak element content than the rest of the CoRoGEE sample and are more abundant towards the inner Galactic disc regions. Almost all of the young [$\alpha$/Fe]-rich stars we
discovered have guiding radii $R_g \leq$ 6 kpc.
Therefore, we tentatively
suggest that the origin of these stars is related to the complex chemical evolution that takes
place near the corotation region of the Galactic bar. 

Unfortunately, some ambiguity remains because the inner Galactic regions accessible to CoRoT are above $|Z_{\mathrm{Gal}}|=0.3$ kpc. This situation is expected to improve by combining 
future APOGEE-2 data with {\it Kepler} seismology 
from the K2 Campaign \citep{Howell2014}, a goal for SDSS-IV.
Further into the future, more information will be obtained from Gaia and the PLATO-2 mission \citep{Rauer2014}, both complemented by spectroscopy for example with the 4MOST facility \citep{deJong2014}.

In a companion paper \citep{Martig2015}, we report the discovery of young-$[\alpha$/Fe]-rich stars in the {\it Kepler} field (although in smaller numbers). Finally, in an ongoing Gaia-ESO follow-up of the CoRoT inner-field stars, more of these stars are found (Valentini et al., in prep.),
providing better statistics and complementing the results shown in this Letter.

\begin{acknowledgements}
The CoRoT space mission, launched on December 27 2006, was developed and 
operated by CNES, with the contribution of Austria, Belgium, Brazil, ESA (RSSD 
and Science Program), Germany and Spain. CC thanks A. Baglin, J. Storm and G. Cescutti for helpful discussions. T.S.R.\ acknowledges support from 
CNPq-Brazil. LG acknowledges support from PRIN INAF 2014.
TM acknowledges financial support from Belspo for contract PRODEX GAIA-DPAC. SM 
acknowledges the support of the NASA grant NNX12AE17G. TCB acknowledges partial support from grants PHY 08-22648; 
Physics Frontier Center/{}JINA, and PHY 14-30152; Physics Frontier Center/{}JINA
Center for the Evolution of the Elements (JINA-CEE), awarded by the US
National Science Foundation. 
The research leading to the presented results has received funding from the European Research 
Council under the European Community's Seventh Framework Programme (FP7/2007-2013) / ERC grant agreement no 338251 (StellarAges).
Funding for the SDSS-III Brazilian Participation Group has been provided by the 
Minist\'erio de Ci\^encia e Tecnologia (MCT), Funda\c{c}\~{a}o Carlos Chagas Filho 
de Amparo \`a Pesquisa do Estado do Rio de Janeiro (FAPERJ), Conselho Nacional de 
Desenvolvimento Cient\'\i fico e Tecnol\'ogico (CNPq), and Financiadora de Estudos e 
Projetos (FINEP).
Funding for SDSS-III has been provided by the Alfred P. Sloan Foundation, the 
Participating Institutions, the National Science Foundation, and the U.S. 
Department of Energy Office of Science. The SDSS-III web site is 
{http://www.sdss3.org/}.
SDSS-III is managed by the Astrophysical Research Consortium for the 
Participating Institutions of the SDSS-III Collaboration including the 
University of Arizona, the Brazilian Participation Group, Brookhaven National 
Laboratory, Carnegie Mellon University, University of Florida, the French 
Participation Group, the German Participation Group, Harvard University, the 
Instituto de Astrofisica de Canarias, the Michigan State/Notre Dame/JINA 
Participation Group, Johns Hopkins University, Lawrence Berkeley National 
Laboratory, Max Planck Institute for Astrophysics, Max Planck Institute for 
Extraterrestrial Physics, New Mexico State University, New York University, 
Ohio State University, Pennsylvania State University, University of Portsmouth, 
Princeton University, the Spanish Participation Group, University of Tokyo, 
University of Utah, Vanderbilt University, University of Virginia, University 
of Washington, and Yale University.
\end{acknowledgements}

\bibliographystyle{aa}
\bibliography{FA_library}

\begin{thebibliography}{42}
\expandafter\ifx\csname natexlab\endcsname\relax\def\natexlab#1{#1}\fi

\bibitem[{{Alam} {et~al.}(2015){Alam}, {Albareti}, {Allende Prieto}, {Anders},
  {Anderson}, {Andrews}, {Armengaud}, {Aubourg}, {Bailey}, {Bautista},
  {Beaton}, {Beers}, {Bender}, {Berlind}, {Beutler}, {Bhardwaj}, {Bird},
  {Bizyaev}, {Blake}, {Blanton}, {Blomqvist}, {Bochanski}, {Bolton}, {Bovy},
  {Shelden Bradley}, {Brandt}, {Brauer}, {Brinkmann}, {Brown}, {Brownstein},
  {Burden}, {Burtin}, {Busca}, {Cai}, {Capozzi}, {Carnero Rosell}, {Carrera},
  {Chen}, {Chiappini}, {Chojnowski}, {Chuang}, {Clerc}, {Comparat}, {Covey},
  {Croft}, {Cuesta}, {Cunha}, {da Costa}, {Da Rio}, {Davenport}, {Dawson}, {De
  Lee}, {Delubac}, {Deshpande}, {Dutra-Ferreira}, {Dwelly}, {Ealet}, {Ebelke},
  {Edmondson}, {Eisenstein}, {Escoffier}, {Esposito}, {Fan},
  {Fern{\'a}ndez-Alvar}, {Feuillet}, {Filiz Ak}, {Finley}, {Finoguenov},
  {Flaherty}, {Fleming}, {Font-Ribera}, {Foster}, {Frinchaboy},
  {Galbraith-Frew}, {Garc{\'{\i}}a-Hern{\'a}ndez}, {Garc{\'{\i}}a P{\'e}rez},
  {Gaulme}, {Ge}, {G{\'e}nova-Santos}, {Ghezzi}, {Gillespie}, {Girardi},
  {Goddard}, {Gontcho}, {Gonz{\'a}lez Hern{\'a}ndez}, {Grebel}, {Niklas Grieb},
  {Grieves}, {Gunn}, {Guo}, {Harding}, {Hasselquist}, {Hawley}, {Hayden},
  {Hearty}, {Ho}, {Hogg}, {Holley-Bockelmann}, {Holtzman}, {Honscheid},
  {Huehnerhoff}, {Jiang}, {Johnson}, {Kinemuchi}, {Kirkby}, {Kitaura},
  {Klaene}, {Kneib}, {Koenig}, {Lam}, {Lan}, {Lang}, {Laurent}, {Le Goff},
  {Leauthaud}, {Lee}, {Lee}, {Licquia}, {Liu}, {Long}, {L{\'o}pez-Corredoira},
  {Lorenzo-Oliveira}, {Lucatello}, {Lundgren}, {Lupton}, {Mack}, {Mahadevan},
  {Maia}, {Majewski}, {Malanushenko}, {Malanushenko}, {Manchado}, {Manera},
  {Mao}, {Maraston}, {Marchwinski}, {Margala}, {Martell}, {Martig}, {Masters},
  {McBride}, {McGehee}, {McGreer}, {McMahon}, {M{\'e}nard}, {Menzel},
  {Merloni}, {M{\'e}sz{\'a}ros}, {Miller}, {Miralda-Escud{\'e}}, {Miyatake},
  {Montero-Dorta}, {More}, {Morice-Atkinson}, {Morrison}, {Muna}, {Myers},
  {Newman}, {Neyrinck}, {Nguyen}, {Nichol}, {Nidever}, {Noterdaeme}, {Nuza},
  {O'Connell}, {O'Connell}, {O'Connell}, {Ogando}, {Olmstead}, {Oravetz},
  {Oravetz}, {Osumi}, {Owen}, {Padgett}, {Padmanabhan}, {Paegert},
  {Palanque-Delabrouille}, {Pan}, {Parejko}, {Park}, {P{\^a}ris},
  {Pattarakijwanich}, {Pellejero-Ibanez}, {Pepper}, {Percival},
  {P{\'e}rez-Fournon}, {P{\'e}rez-R{\`a}fols}, {Petitjean}, {Pieri},
  {Pinsonneault}, {Porto de Mello}, {Prada}, {Prakash}, {Price-Whelan},
  {Raddick}, {Rahman}, {Reid}, {Rich}, {Rix}, {Robin}, {Rockosi}, {Rodrigues},
  {Rodr{\'{\i}}guez-Rottes}, {Roe}, {Ross}, {Ross}, {Rossi}, {Ruan},
  {Rubi{\~n}o-Mart$\backslash$'$\backslash$in}, {Rykoff}, {Salazar-Albornoz},
  {Salvato}, {Samushia}, {S{\'a}nchez}, {Santiago}, {Sayres}, {Schiavon},
  {Schlegel}, {Schmidt}, {Schneider}, {Schultheis}, {Schwope}, {Sc{\'o}ccola},
  {Sellgren}, {Seo}, {Shane}, {Shen}, {Shetrone}, {Shu}, {Sivarani},
  {Skrutskie}, {Slosar}, {Smith}, {Sobreira}, {Stassun}, {Steinmetz},
  {Strauss}, {Streblyanska}, {Swanson}, {Tan}, {Tayar}, {Terrien}, {Thakar},
  {Thomas}, {Thompson}, {Tinker}, {Tojeiro}, {Troup}, {Vargas-Maga{\~n}a},
  {Vazquez}, {Verde}, {Viel}, {Vogt}, {Wake}, {Wang}, {Weaver}, {Weinberg},
  {Weiner}, {White}, {Wilson}, {Wisniewski}, {Wood-Vasey}, {Y{\`e}che}, {York},
  {Zakamska}, {Zamora}, {Zasowski}, {Zehavi}, {Zhao}, {Zheng}, {Zhou}, {Zhou},
  {Zhu}, \& {Zou}}]{Alam2015}
{Alam}, S., {Albareti}, F.~D., {Allende Prieto}, C., {et~al.} 2015, ArXiv
  e-prints:1501.00963

\bibitem[{{Anders} {et~al.}(2014){Anders}, {Chiappini}, {Santiago},
  {Rocha-Pinto}, {Girardi}, {da Costa}, {Maia}, {Steinmetz}, {Minchev},
  {Schultheis}, {Boeche}, {Miglio}, {Montalb{\'a}n}, {Schneider}, {Beers},
  {Cunha}, {Allende Prieto}, {Balbinot}, {Bizyaev}, {Brauer}, {Brinkmann},
  {Frinchaboy}, {Garc{\'{\i}}a P{\'e}rez}, {Hayden}, {Hearty}, {Holtzman},
  {Johnson}, {Kinemuchi}, {Majewski}, {Malanushenko}, {Malanushenko},
  {Nidever}, {O'Connell}, {Pan}, {Robin}, {Schiavon}, {Shetrone}, {Skrutskie},
  {Smith}, {Stassun}, \& {Zasowski}}]{Anders2014}
{Anders}, F., {Chiappini}, C., {Santiago}, B.~X., {et~al.} 2014, \aap, 564,
  A115

\bibitem[{{Baglin} {et~al.}(2006){Baglin}, {Auvergne}, {Barge}, {Deleuil},
  {Catala}, {Michel}, {Weiss}, \& {COROT Team}}]{Baglin2006}
{Baglin}, A., {Auvergne}, M., {Barge}, P., {et~al.} 2006, in ESA Special
  Publication, ed. M.~{Fridlund}, A.~{Baglin}, J.~{Lochard}, \& L.~{Conroy},
  Vol. 1306, 33

\bibitem[{{Bensby} {et~al.}(2014){Bensby}, {Feltzing}, \& {Oey}}]{Bensby2014}
{Bensby}, T., {Feltzing}, S., \& {Oey}, M.~S. 2014, \aap, 562, A71

\bibitem[{{Bergemann} {et~al.}(2014){Bergemann}, {Ruchti}, {Serenelli},
  {Feltzing}, {Alves-Brito}, {Asplund}, {Bensby}, {Gruyters}, {Heiter},
  {Hourihane}, {Korn}, {Lind}, {Marino}, {Jofre}, {Nordlander}, {Ryde},
  {Worley}, {Gilmore}, {Randich}, {Ferguson}, {Jeffries}, {Micela},
  {Negueruela}, {Prusti}, {Rix}, {Vallenari}, {Alfaro}, {Allende Prieto},
  {Bragaglia}, {Koposov}, {Lanzafame}, {Pancino}, {Recio-Blanco}, {Smiljanic},
  {Walton}, {Costado}, {Franciosini}, {Hill}, {Lardo}, {de Laverny}, {Magrini},
  {Maiorca}, {Masseron}, {Morbidelli}, {Sacco}, {Kordopatis}, \& {Tautvai{\v
  s}ien{\.e}}}]{Bergemann2014}
{Bergemann}, M., {Ruchti}, G.~R., {Serenelli}, A., {et~al.} 2014, \aap, 565,
  A89

\bibitem[{{Bissantz} {et~al.}(2003){Bissantz}, {Englmaier}, \&
  {Gerhard}}]{Bissantz2003}
{Bissantz}, N., {Englmaier}, P., \& {Gerhard}, O. 2003, \mnras, 340, 949

\bibitem[{{Chiappini}(2009)}]{Chiappini2009}
{Chiappini}, C. 2009, in IAU Symposium, Vol. 254, IAU Symposium, ed.
  J.~{Andersen}, B.~{Nordstr{\"o}m}, \& J.~{Bland-Hawthorn}, 191--196

\bibitem[{{Chiappini} {et~al.}(1997){Chiappini}, {Matteucci}, \&
  {Gratton}}]{Chiappini1997}
{Chiappini}, C., {Matteucci}, F., \& {Gratton}, R. 1997, \apj, 477, 765

\bibitem[{{Combes}(2014)}]{Combes2014}
{Combes}, F. 2014, in ASP Conference Series, Vol. 480, Structure and Dynamics
  of Disk Galaxies, ed. M.~S. {Seigar} \& P.~{Treuthardt}, 211

\bibitem[{{Cunha} {et~al.}(2007){Cunha}, {Sellgren}, {Smith}, {Ramirez},
  {Blum}, \& {Terndrup}}]{Cunha2007}
{Cunha}, K., {Sellgren}, K., {Smith}, V.~V., {et~al.} 2007, \apj, 669, 1011

\bibitem[{{da Silva} {et~al.}(2006){da Silva}, {Girardi}, {Pasquini},
  {Setiawan}, {von der L{\"u}he}, {de Medeiros}, {Hatzes}, {D{\"o}llinger}, \&
  {Weiss}}]{daSilva2006}
{da Silva}, L., {Girardi}, L., {Pasquini}, L., {et~al.} 2006, \aap, 458, 609

\bibitem[{{Davies} {et~al.}(2009{\natexlab{a}}){Davies}, {Origlia},
  {Kudritzki}, {Figer}, {Rich}, \& {Najarro}}]{Davies2009}
{Davies}, B., {Origlia}, L., {Kudritzki}, R.-P., {et~al.} 2009{\natexlab{a}},
  \apj, 694, 46

\bibitem[{{Davies} {et~al.}(2009{\natexlab{b}}){Davies}, {Origlia},
  {Kudritzki}, {Figer}, {Rich}, {Najarro}, {Negueruela}, \&
  {Clark}}]{Davies2009a}
{Davies}, B., {Origlia}, L., {Kudritzki}, R.-P., {et~al.} 2009{\natexlab{b}},
  \apj, 696, 2014

\bibitem[{{de Jong} {et~al.}(2014){de Jong}, {Barden}, {Bellido-Tirado},
  {Brynnel}, {Chiappini}, {Depagne}, {Haynes}, {Johl}, {Phillips}, {Schnurr},
  {Schwope}, {Walcher}, {Bauer}, {Cescutti}, {Cioni}, {Dionies}, {Enke},
  {Haynes}, {Kelz}, {Kitaura}, {Lamer}, {Minchev}, {M{\"u}ller}, {Nuza},
  {Olaya}, {Piffl}, {Popow}, {Saviauk}, {Steinmetz}, {Ural}, {Valentini},
  {Winkler}, {Wisotzki}, {Ansorge}, {Banerji}, {Gonzalez Solares}, {Irwin},
  {Kennicutt}, {King}, {McMahon}, {Koposov}, {Parry}, {Sun}, {Walton},
  {Finger}, {Iwert}, {Krumpe}, {Lizon}, {Mainieri}, {Amans}, {Bonifacio},
  {Cohen}, {Fran{\c c}ois}, {Jagourel}, {Mignot}, {Royer}, {Sartoretti},
  {Bender}, {Hess}, {Lang-Bardl}, {Muschielok}, {Schlichter}, {B{\"o}hringer},
  {Boller}, {Bongiorno}, {Brusa}, {Dwelly}, {Merloni}, {Nandra}, {Salvato},
  {Pragt}, {Navarro}, {Gerlofsma}, {Roelfsema}, {Dalton}, {Middleton}, {Tosh},
  {Boeche}, {Caffau}, {Christlieb}, {Grebel}, {Hansen}, {Koch}, {Ludwig},
  {Mandel}, {Quirrenbach}, {Sbordone}, {Seifert}, {Thimm}, {Helmi}, {trager},
  {Bensby}, {Feltzing}, {Ruchti}, {Edvardsson}, {Korn}, {Lind}, {Boland},
  {Colless}, {Frost}, {Gilbert}, {Gillingham}, {Lawrence}, {Legg}, {Saunders},
  {Sheinis}, {Driver}, {Robotham}, {Bacon}, {Caillier}, {Kosmalski}, {Laurent},
  \& {Richard}}]{deJong2014}
{de Jong}, R.~S., {Barden}, S., {Bellido-Tirado}, O., {et~al.} 2014, in SPIE
  Conference Series, Vol. 9147, SPIE Conference Series

\bibitem[{{Eisenstein} {et~al.}(2011){Eisenstein}, {Weinberg}, {Agol},
  {Aihara}, {Allende Prieto}, {Anderson}, {Arns}, {Aubourg}, {Bailey},
  {Balbinot}, \& et~al.}]{Eisenstein2011}
{Eisenstein}, D.~J., {Weinberg}, D.~H., {Agol}, E., {et~al.} 2011, \aj, 142, 72

\bibitem[{{Feast} {et~al.}(2014){Feast}, {Menzies}, {Matsunaga}, \&
  {Whitelock}}]{Feast2014}
{Feast}, M.~W., {Menzies}, J.~W., {Matsunaga}, N., \& {Whitelock}, P.~A. 2014,
  \nat, 509, 342

\bibitem[{{Fuhrmann}(2011)}]{Fuhrmann2011}
{Fuhrmann}, K. 2011, \mnras, 414, 2893

\bibitem[{{Genovali} {et~al.}(2014){Genovali}, {Lemasle}, {Bono}, {Romaniello},
  {Fabrizio}, {Ferraro}, {Iannicola}, {Laney}, {Nonino}, {Bergemann},
  {Buonanno}, {Fran{\c c}ois}, {Inno}, {Kudritzki}, {Matsunaga}, {Pedicelli},
  {Primas}, \& {Th{\'e}venin}}]{Genovali2014}
{Genovali}, K., {Lemasle}, B., {Bono}, G., {et~al.} 2014, \aap, 566, A37

\bibitem[{{Gilliland} {et~al.}(2010){Gilliland}, {Brown},
  {Christensen-Dalsgaard}, {Kjeldsen}, {Aerts}, {Appourchaux}, {Basu},
  {Bedding}, {Chaplin}, {Cunha}, {De Cat}, {De Ridder}, {Guzik}, {Handler},
  {Kawaler}, {Kiss}, {Kolenberg}, {Kurtz}, {Metcalfe}, {Monteiro}, {Szab{\'o}},
  {Arentoft}, {Balona}, {Debosscher}, {Elsworth}, {Quirion}, {Stello},
  {Su{\'a}rez}, {Borucki}, {Jenkins}, {Koch}, {Kondo}, {Latham}, {Rowe}, \&
  {Steffen}}]{Gilliland2010}
{Gilliland}, R.~L., {Brown}, T.~M., {Christensen-Dalsgaard}, J., {et~al.} 2010,
  \pasp, 122, 131

\bibitem[{{Gunn} {et~al.}(2006){Gunn}, {Siegmund}, {Mannery}, {Owen}, {Hull},
  {Leger}, {Carey}, {Knapp}, {York}, {Boroski}, {Kent}, {Lupton}, {Rockosi},
  {Evans}, {Waddell}, {Anderson}, {Annis}, {Barentine}, {Bartoszek}, {Bastian},
  {Bracker}, {Brewington}, {Briegel}, {Brinkmann}, {Brown}, {Carr},
  {Czarapata}, {Drennan}, {Dombeck}, {Federwitz}, {Gillespie}, {Gonzales},
  {Hansen}, {Harvanek}, {Hayes}, {Jordan}, {Kinney}, {Klaene}, {Kleinman},
  {Kron}, {Kresinski}, {Lee}, {Limmongkol}, {Lindenmeyer}, {Long}, {Loomis},
  {McGehee}, {Mantsch}, {Neilsen}, {Neswold}, {Newman}, {Nitta}, {Peoples},
  {Pier}, {Prieto}, {Prosapio}, {Rivetta}, {Schneider}, {Snedden}, \&
  {Wang}}]{Gunn2006}
{Gunn}, J.~E., {Siegmund}, W.~A., {Mannery}, E.~J., {et~al.} 2006, \aj, 131,
  2332

\bibitem[{{Haywood} {et~al.}(2013){Haywood}, {Di Matteo}, {Lehnert}, {Katz}, \&
  {G{\'o}mez}}]{Haywood2013}
{Haywood}, M., {Di Matteo}, P., {Lehnert}, M.~D., {Katz}, D., \& {G{\'o}mez},
  A. 2013, \aap, 560, A109

\bibitem[{{Holtzman} {et~al.}(2015){Holtzman}, {Shetrone}, {Johnson}, {Allende
  Prieto}, {Anders}, {Andrews}, {Beers}, {Bizyaev}, {Blanton}, {Bovy},
  {Carrera}, {Cunha}, {Eisenstein}, {Feuillet}, {Frinchaboy}, {Galbraith-Frew},
  {Garcia Perez}, {Anibal Garcia Hernandez}, {Hasselquist}, {Hayden}, {Hearty},
  {Ivans}, {Majewski}, {Martell}, {Meszaros}, {Muna}, {Nidever}, {Nguyen},
  {O'Connell}, {Pan}, {Pinsonneault}, {Robin}, {Schiavon}, {Shane}, {Sobeck},
  {Smith}, {Troup}, {Weinberg}, {Wilson}, {Wood-Vasey}, {Zamora}, \&
  {Zasowski}}]{Holtzman2015}
{Holtzman}, J.~A., {Shetrone}, M., {Johnson}, J.~A., {et~al.} 2015, ArXiv
  e-prints:1501.04110

\bibitem[{{Howell} {et~al.}(2014){Howell}, {Sobeck}, {Haas}, {Still},
  {Barclay}, {Mullally}, {Troeltzsch}, {Aigrain}, {Bryson}, {Caldwell},
  {Chaplin}, {Cochran}, {Huber}, {Marcy}, {Miglio}, {Najita}, {Smith},
  {Twicken}, \& {Fortney}}]{Howell2014}
{Howell}, S.~B., {Sobeck}, C., {Haas}, M., {et~al.} 2014, \pasp, 126, 398

\bibitem[{{Lebreton} \& {Goupil}(2014)}]{Lebreton2014a}
{Lebreton}, Y. \& {Goupil}, M.-J. 2014, ArXiv e-prints:1406.0652

\bibitem[{{Lebreton} {et~al.}(2014){Lebreton}, {Goupil}, \&
  {Montalb{\'a}n}}]{Lebreton2014}
{Lebreton}, Y., {Goupil}, M.~J., \& {Montalb{\'a}n}, J. 2014, in EAS
  Publications Series, Vol.~65, EAS Publications Series, 99--176

\bibitem[{{Luck} \& {Lambert}(2011)}]{Luck2011a}
{Luck}, R.~E. \& {Lambert}, D.~L. 2011, \aj, 142, 136

\bibitem[{{Martig} {et~al.}(2015){Martig}, {Rix}, {Silva Aguirre}, {Hekker},
  {Mosser}, {Elsworth}, {Bovy}, {Stello}, {Anders}, {Garc{\'{\i}}a}, {Tayar},
  {Rodrigues}, {Basu}, {Carrera}, {Ceillier}, {Chaplin}, {Chiappini},
  {Frinchaboy}, {Garc{\'{\i}}a-Hern{\'a}ndez}, {Hearty}, {Holtzman}, {Johnson},
  {Mathur}, {M{\'e}sz{\'a}ros}, {Miglio}, {Nidever}, {Pinsonneault},
  {Schiavon}, {Schneider}, {Serenelli}, {Shetrone}, \& {Zamora}}]{Martig2015}
{Martig}, M., {Rix}, H.-W., {Silva Aguirre}, V., {et~al.} 2015, ArXiv
  e-prints:1412.3453

\bibitem[{{Matteucci}(2001)}]{Matteucci2001}
{Matteucci}, F., ed. 2001, Astrophysics and Space Science Library, Vol. 253,
  {The chemical evolution of the Galaxy}

\bibitem[{{M{\'e}sz{\'a}ros} {et~al.}(2013){M{\'e}sz{\'a}ros}, {Holtzman},
  {Garc{\'{\i}}a P{\'e}rez}, {Allende Prieto}, {Schiavon}, {Basu}, {Bizyaev},
  {Chaplin}, {Chojnowski}, {Cunha}, {Elsworth}, {Epstein}, {Frinchaboy},
  {Garc{\'{\i}}a}, {Hearty}, {Hekker}, {Johnson}, {Kallinger}, {Koesterke},
  {Majewski}, {Martell}, {Nidever}, {Pinsonneault}, {O'Connell}, {Shetrone},
  {Smith}, {Wilson}, \& {Zasowski}}]{Meszaros2013}
{M{\'e}sz{\'a}ros}, S., {Holtzman}, J., {Garc{\'{\i}}a P{\'e}rez}, A.~E.,
  {et~al.} 2013, \aj, 146, 133

\bibitem[{{Miglio} {et~al.}(2013b){Miglio}, {Chiappini}, {Morel}, {Barbieri},
  {Chaplin}, {Girardi}, {Montalb{\'a}n}, {Noels}, {Valentini}, {Mosser},
  {Baudin}, {Casagrande}, {Fossati}, {Silva Aguirre}, \&
  {Baglin}}]{Miglio2013b}
{Miglio}, A., {Chiappini}, C., {Morel}, T., {et~al.} 2013b, in European
  Physical Journal Web of Conferences, Vol.~43, 3004

\bibitem[{{Miglio} {et~al.}(2013a){Miglio}, {Chiappini}, {Morel}, {Barbieri},
  {Chaplin}, {Girardi}, {Montalb{\'a}n}, {Valentini}, {Mosser}, {Baudin},
  {Casagrande}, {Fossati}, {Aguirre}, \& {Baglin}}]{Miglio2013a}
{Miglio}, A., {Chiappini}, C., {Morel}, T., {et~al.} 2013a, \mnras, 429, 423

\bibitem[{{Minchev} {et~al.}(2013){Minchev}, {Chiappini}, \&
  {Martig}}]{Minchev2013}
{Minchev}, I., {Chiappini}, C., \& {Martig}, M. 2013, \aap, 558, A9

\bibitem[{{Minchev} {et~al.}(2014){Minchev}, {Chiappini}, \&
  {Martig}}]{Minchev2014}
{Minchev}, I., {Chiappini}, C., \& {Martig}, M. 2014, \aap, 572, A92

\bibitem[{{Mosser} \& {Appourchaux}(2009)}]{Mosser2009}
{Mosser}, B. \& {Appourchaux}, T. 2009, \aap, 508, 877

\bibitem[{{Mosser} {et~al.}(2011){Mosser}, {Barban}, {Montalb{\'a}n}, {Beck},
  {Miglio}, {Belkacem}, {Goupil}, {Hekker}, {De Ridder}, {Dupret}, {Elsworth},
  {Noels}, {Baudin}, {Michel}, {Samadi}, {Auvergne}, {Baglin}, \&
  {Catala}}]{Mosser2011}
{Mosser}, B., {Barban}, C., {Montalb{\'a}n}, J., {et~al.} 2011, \aap, 532, A86

\bibitem[{{Mosser} {et~al.}(2010){Mosser}, {Belkacem}, {Goupil}, {Miglio},
  {Morel}, {Barban}, {Baudin}, {Hekker}, {Samadi}, {De Ridder}, {Weiss},
  {Auvergne}, \& {Baglin}}]{Mosser2010}
{Mosser}, B., {Belkacem}, K., {Goupil}, M.-J., {et~al.} 2010, \aap, 517, A22

\bibitem[{{Origlia} {et~al.}(2013){Origlia}, {Oliva}, {Maiolino},
  {Mucciarelli}, {Baffa}, {Biliotti}, {Bruno}, {Falcini}, {Gavriousev},
  {Ghinassi}, {Giani}, {Gonzalez}, {Leone}, {Lodi}, {Massi}, {Montegriffo},
  {Mochi}, {Pedani}, {Rossetti}, {Scuderi}, {Sozzi}, \& {Tozzi}}]{Origlia2013}
{Origlia}, L., {Oliva}, E., {Maiolino}, R., {et~al.} 2013, \aap, 560, A46

\bibitem[{{Pagel}(2009)}]{Pagel2009}
{Pagel}, B.~E.~J. 2009, {Nucleosynthesis and Chemical Evolution of Galaxies}

\bibitem[{{Ram{\'{\i}}rez} {et~al.}(2007){Ram{\'{\i}}rez}, {Allende Prieto}, \&
  {Lambert}}]{Ramirez2007}
{Ram{\'{\i}}rez}, I., {Allende Prieto}, C., \& {Lambert}, D.~L. 2007, \aap,
  465, 271

\bibitem[{{Rauer} {et~al.}(2014){Rauer}, {Catala}, {Aerts}, {Appourchaux},
  {Benz}, {Brandeker}, {Christensen-Dalsgaard}, {Deleuil}, {Gizon}, {Goupil},
  {G{\"u}del}, {Janot-Pacheco}, {Mas-Hesse}, {Pagano}, {Piotto}, {Pollacco},
  {Santos}, {Smith}, {Su{\'a}rez}, {Szab{\'o}}, {Udry}, {Adibekyan}, {Alibert},
  {Almenara}, {Amaro-Seoane}, {Eiff}, {Asplund}, {Antonello}, {Barnes},
  {Baudin}, {Belkacem}, {Bergemann}, {Bihain}, {Birch}, {Bonfils}, {Boisse},
  {Bonomo}, {Borsa}, {Brand{\~a}o}, {Brocato}, {Brun}, {Burleigh}, {Burston},
  {Cabrera}, {Cassisi}, {Chaplin}, {Charpinet}, {Chiappini}, {Church},
  {Csizmadia}, {Cunha}, {Damasso}, {Davies}, {Deeg}, {D{\'{\i}}az}, {Dreizler},
  {Dreyer}, {Eggenberger}, {Ehrenreich}, {Eigm{\"u}ller}, {Erikson}, {Farmer},
  {Feltzing}, {de Oliveira Fialho}, {Figueira}, {Forveille}, {Fridlund},
  {Garc{\'{\i}}a}, {Giommi}, {Giuffrida}, {Godolt}, {Gomes da Silva},
  {Granzer}, {Grenfell}, {Grotsch-Noels}, {G{\"u}nther}, {Haswell}, {Hatzes},
  {H{\'e}brard}, {Hekker}, {Helled}, {Heng}, {Jenkins}, {Johansen},
  {Khodachenko}, {Kislyakova}, {Kley}, {Kolb}, {Krivova}, {Kupka}, {Lammer},
  {Lanza}, {Lebreton}, {Magrin}, {Marcos-Arenal}, {Marrese}, {Marques},
  {Martins}, {Mathis}, {Mathur}, {Messina}, {Miglio}, {Montalban}, {Montalto},
  {Monteiro}, {Moradi}, {Moravveji}, {Mordasini}, {Morel}, {Mortier},
  {Nascimbeni}, {Nelson}, {Nielsen}, {Noack}, {Norton}, {Ofir}, {Oshagh},
  {Ouazzani}, {P{\'a}pics}, {Parro}, {Petit}, {Plez}, {Poretti}, {Quirrenbach},
  {Ragazzoni}, {Raimondo}, {Rainer}, {Reese}, {Redmer}, {Reffert},
  {Rojas-Ayala}, {Roxburgh}, {Salmon}, {Santerne}, {Schneider}, {Schou},
  {Schuh}, {Schunker}, {Silva-Valio}, {Silvotti}, {Skillen}, {Snellen}, {Sohl},
  {Sousa}, {Sozzetti}, {Stello}, {Strassmeier}, {{\v S}vanda}, {Szab{\'o}},
  {Tkachenko}, {Valencia}, {Van Grootel}, {Vauclair}, {Ventura}, {Wagner},
  {Walton}, {Weingrill}, {Werner}, {Wheatley}, \& {Zwintz}}]{Rauer2014}
{Rauer}, H., {Catala}, C., {Aerts}, C., {et~al.} 2014, Experimental Astronomy,
  38, 249

\bibitem[{{Rodrigues} {et~al.}(2014){Rodrigues}, {Girardi}, {Miglio},
  {Bossini}, {Bovy}, {Epstein}, {Pinsonneault}, {Stello}, {Zasowski}, {Prieto},
  {Chaplin}, {Hekker}, {Johnson}, {M{\'e}sz{\'a}ros}, {Mosser}, {Anders},
  {Basu}, {Beers}, {Chiappini}, {da Costa}, {Elsworth}, {Garc{\'{\i}}a},
  {P{\'e}rez}, {Hearty}, {Maia}, {Majewski}, {Mathur}, {Montalb{\'a}n},
  {Nidever}, {Santiago}, {Schultheis}, {Serenelli}, \&
  {Shetrone}}]{Rodrigues2014}
{Rodrigues}, T.~S., {Girardi}, L., {Miglio}, A., {et~al.} 2014, \mnras, 445,
  2758

\bibitem[{{Yong} {et~al.}(2012){Yong}, {Carney}, \& {Friel}}]{Yong2012}
{Yong}, D., {Carney}, B.~W., \& {Friel}, E.~D. 2012, \aj, 144, 95

\end{thebibliography}

\newpage
\begin{appendix}

\section{Best-candidate young $\alpha$-enhanced stars}

Table~\ref{A1} summarises our measured quantities for 
the best-candidate young $\alpha$-enhanced stars (17 $2\sigma$-outliers; large pentagons in Fig. \ref{Fig1}, and 11 
1$\sigma$-outliers; stars in Fig. \ref{Fig1}).
We first report our input values: the adopted seismic parameters $\Delta \nu$ and $\nu_{\mathrm{max}}$ (as computed by 
automatic as well as supervised analyses of the CoRoT light curves),
ASPCAP spectroscopic parameters $T_{\mathrm{eff}}$, [Fe/H], [$\alpha$/Fe], and the number of APOGEE observations, ${\mathrm{N_{APO}}}$.
We note that all stars in question have been observed at very high signal-to-noise ratios ($S/N>140$ per resolution element). The radial-velocity scatter between subsequent
observations is always smaller than 0.6 km/s; meaning that their values are consistent with all stars being single stars or widely separated binaries.

We also present the estimated stellar masses $\mathrm{M_{scale}}$, as determined from seismic scaling relations and the 1$\sigma$ upper-limits for the ages 
(as determined by PARAM). A comparison of the masses estimated by PARAM and those inferred directly from the scaling
relations is reported in A15 for the full CoRoGEE sample.
Also listed are the current Galactocentric positions $R_{\mathrm{Gal}}$ and $Z_{\mathrm{Gal}}$ and the
guiding radius $R_g$ of each star. 

We also show a note on the quality of the light curves ($Q$) and 
a flag based on the supervised analysis. 
Because the automated and supervised analyses sometimes yield different results, we recomputed masses and ages using the individually obtained $\Delta\nu$ and $\nu_{\mathrm{max}}$ values and updated uncertainties where necessary. As expected, the numbers of the young $\alpha$-enhanced stars are slightly different. In Table \ref{A1}, we only report the robust $2\sigma-$ and $1\sigma-$outliers.

However, these uncertainties have a small impact on our main result, as after the individual analysis, still 17 stars out of 20 seem to be younger than 4 Gyr.
Individual supervised analysis shows that:

\begin{enumerate}
\item One star (CoRoT 101071033) had to be excluded from the parent sample due to the very poor quality of its light curve;
\item Four stars that seemed to be $2\sigma$-outliers were shifted to older ages:
CoRoT 101093867, 101071033, 102645343, and 10264381. Similarly, seven candidate $1\sigma$-outliers fall out of the sample: CoRoT 101057962, 101041814, 102626343, 100886873, 101208801, 101212022, and 101227666.
\item CoRoT 101093867 is a complex case, where both $\Delta \nu$ values appear as possible solutions; for six other stars, another solution is possible, because the 
lightcurve SNR is not high enough to undoubtedly resolve the radial/dipole mode possible mismatch (such cases cannot be seen in the general blind automated analysis);
\item For CoRoT 100958571, the solution obtained through supervised fitting, close to the one found by the automated pipeline, should be preferred. Also, for most of the remaining stars, slight improvements in the determination of the seismic parameters are possible.
\end{enumerate}


\begin{sidewaystable*}
\vspace{2cm}
\caption{Best-candidate young $\alpha$-enhanced stars: seismic and 
spectroscopic adopted parameters and uncertainties, stellar 
masses and ages, current Galactocentric positions
${R_{\mathrm{Gal}}}$ and ${Z_{\mathrm{Gal}}}$, and guiding-centre radii $R_{\mathrm{g}}$.}\label{A1}
{\centering
\tiny
\setlength{\tabcolsep}{5pt}
\begin{tabular}{rlcccrrcrcrrrrrrrrr}
\hline
\hline
  \multicolumn{1}{c}{CoRoT ID} &
  \multicolumn{1}{c}{APOGEE ID} &
  \multicolumn{1}{c}{$\Delta \nu$} &
  \multicolumn{1}{c}{$\nu_{\mathrm{max}}$}&
  \multicolumn{1}{c}{$Q^a$}&
  \multicolumn{1}{c}{${\Delta \nu_i}^b$}&  
  \multicolumn{1}{c}{${\nu_{\mathrm{max}_i}}^b$} &  
  \multicolumn{1}{c}{Flag$^c$} &
  \multicolumn{1}{c}{${\mathrm{N_{APO}}}$} &
  \multicolumn{1}{c}{${\mathrm{T_{eff}}}^d$} &
  \multicolumn{1}{c}{${\mathrm{[Fe/H]}}$} &
  \multicolumn{1}{c}{$[\alpha{\mathrm{/Fe]}}$} &
  \multicolumn{1}{c}{$\mathrm{M_{scale}}$} &
  \multicolumn{1}{c}{$\mathrm{\tau_{68U}}^e$} &
  \multicolumn{1}{c}{$\mathrm{\tau_{68U}^{i}}^f$}  &
  \multicolumn{1}{c}{${R_{\mathrm{Gal}}}^g$} &
  \multicolumn{1}{c}{${Z_{\mathrm{Gal}}}^h$} &
  \multicolumn{1}{c}{$R_{\mathrm{g}}$} &
\\
  \multicolumn{1}{c}{} &
  \multicolumn{1}{c}{} &
  \multicolumn{1}{c}{[$\mu$Hz]} &
  \multicolumn{1}{c}{[$\mu$Hz]}&
  \multicolumn{1}{c}{}&
  \multicolumn{1}{c}{[$\mu$Hz]}&  
  \multicolumn{1}{c}{[$\mu$Hz]} &  
  \multicolumn{1}{c}{}&
  \multicolumn{1}{c}{} &
  \multicolumn{1}{c}{[K]} &
  \multicolumn{1}{c}{} &
  \multicolumn{1}{c}{} &
  \multicolumn{1}{c}{[M$_{\odot}$]} &
  \multicolumn{1}{c}{[Gyr]} &
  \multicolumn{1}{c}{[Gyr]}  &
  \multicolumn{1}{c}{[kpc]} &
  \multicolumn{1}{c}{[kpc]} &
  \multicolumn{1}{c}{[kpc]} &
\\
   \hline
\multicolumn{18}{c}{\it \small $2\sigma$-outliers} \\
  \hline
  100580176 & 2M19232036+0116385 &   $ 1.2  \pm  0.01 $ &   $   8.11 \pm  0.22  $ & OK   &    1.27 &      8.0  &      1 &       1 &   4200 & $-0.2  \pm 0.03 $ & $ 0.16 \pm 0.01 $ &    $ 1.49 \pm  0.22 $ &        2.5 &        4.5 &   6.06 &  $-0.29$  & $   9.1 \pm   0.5 $ \\
  100692726 & 2M19240121+0115468 &   $ 2.71 \pm  0.03 $ &   $  22.41 \pm  0.58  $ & OK   &    2.7  &     22.3  &      0 &       7 &   4390 & $-0.11 \pm 0.03 $ & $ 0.16 \pm 0.01 $ &    $ 1.48 \pm  0.14 $ &        4.3 &        4.3 &   4.91 &  $-0.75$  & $   4.2 \pm   1.2 $ \\
  100958571 & 2M19253009+0100237 &   $ 1.94 \pm  0.04 $ &   $  14.72 \pm  0.65  $ & OK   &    1.97 &     14.7  &      2 &       3 &   4410 & $-0.55 \pm 0.04 $ & $ 0.20 \pm 0.02 $ &    $ 1.51 \pm  0.24 $ &        3.4 &        4.7 &   5.46 &  $-0.46$  & $   5.5 \pm   0.9 $ \\
  101045095 & 2M19260245+0003446 &   $ 2.78 \pm  0.04 $ &   $  22.17 \pm  0.64  $ & poor &    2.8  &     22.6  &      0 &       3 &   4400 & $-0.23 \pm 0.03 $ & $ 0.23 \pm 0.01 $ &    $ 1.34 \pm  0.14 $ &        6.0 &        5.7 &   5.87 &  $-0.38$  &                     \\
  101072104 & 2M19261545+0011507 &   $ 3.01 \pm  0.04 $ &   $  23.90 \pm  0.71  $ & OK   &    3.01 &     24.8  &      0 &       7 &   4580 & $-0.42 \pm 0.04 $ & $ 0.24 \pm 0.02 $ &    $ 1.41 \pm  0.14 $ &        5.8 &        3.6 &   4.87 &  $-0.74$  & $   3.7 \pm   1.2 $ \\
  101100354 & 2M19262657+0144163 &   $ 4.56 \pm  0.04 $ &   $  41.60 \pm  0.93  $ & poor &    4.34 &     43.6  &      2 &       7 &   4520 & $-0.12 \pm 0.03 $ & $ 0.21 \pm 0.01 $ &    $ 1.74 \pm  0.14 $ &        7.6 &        3.0 &   5.97 &  $-0.34$  & $   5.3 \pm   0.7 $ \\
  101113416 & 2M19263149+0159448 &   $ 1.11 \pm  0.01 $ &   $   6.79 \pm  0.20  $ & OK   &    1.14 &      6.74 &      1 &       3 &   4360 & $-0.48 \pm 0.04 $ & $ 0.24 \pm 0.02 $ &    $ 1.36 \pm  0.15 $ &        3.6 &        4.5 &   5.14 &  $-0.58$  & $   1.7 \pm   1.2 $ \\
  101114706 & 2M19263197-0035004 &   $ 0.97 \pm  0.02 $ &   $   6.14 \pm  0.31  $ & OK   &    0.98 &      6.14 &      0 &       3 &   4170 & $-0.27 \pm 0.03 $ & $ 0.19 \pm 0.01 $ &    $ 1.65 \pm  0.29 $ &        3.5 &        3.8 &   4.76 &  $-0.84$  & $   2.0 \pm   1.0 $ \\
  101121769 & 2M19263465+0004069 &   $ 1.34 \pm  0.03 $ &   $   8.88 \pm  0.35  $ & OK   &    1.34 &      8.88 &      0 &       3 &   4340 & $-0.34 \pm 0.03 $ & $ 0.17 \pm 0.02 $ &    $ 1.52 \pm  0.23 $ &        4.0 &        4.0 &   5.12 &  $-0.61$  &                     \\
  101138968 & 2M19264111+0214048 &   $ 2.46 \pm  0.04 $ &   $  20.74 \pm  0.73  $ & OK   &    2.46 &     20.7  &      0 &       7 &   4500 & $-0.45 \pm 0.04 $ & $ 0.27 \pm 0.02 $ &    $ 1.79 \pm  0.22 $ &        2.1 &        2.3 &   5.03 &  $-0.72$  & $   3.2 \pm   1.1 $ \\
  101342375 & 2M19280053+0016331 &   $ 2.06 \pm  0.04 $ &   $  16.69 \pm  0.74  $ & OK   &    2.00 &     16.7  &      0 &       7 &   4340 & $ 0.03 \pm 0.03 $ & $ 0.15 \pm 0.01 $ &    $ 2.03 \pm  0.32 $ &        2.6 &        2.4 &   4.86 &  $-0.83$  & $   6.6 \pm   1.2 $ \\
  101386073 & 2M19282189+0010322 &   $ 5.21 \pm  0.07 $ &   $  48.40 \pm  1.41  $ & OK   &    5.23 &     51.2  &      0 &       7 &   4610 & $-0.33 \pm 0.04 $ & $ 0.19 \pm 0.02 $ &    $ 1.37 \pm  0.14 $ &        6.9 &        3.9 &   5.87 &  $-0.41$  &                     \\
  101415638 & 2M19283410+0006205 &   $ 5.21 \pm  0.11 $ &   $  47.68 \pm  2.28  $ & poor &    4.80 &     47.7  &      1 &       7 &   4960 & $-0.53 \pm 0.04 $ & $ 0.22 \pm 0.02 $ &    $ 1.48 \pm  0.32 $ &        2.7 &        2.1 &   5.74 &  $-0.45$  & $   5.5 \pm   0.7 $ \\
  101594554 & 2M19294723+0007020 &   $ 2.70 \pm  0.03 $ &   $  21.52 \pm  0.51  $ & OK   &    2.72 &     21.73 &      0 &       7 &   4430 & $-0.29 \pm 0.04 $ & $ 0.17 \pm 0.01 $ &    $ 1.35 \pm  0.11 $ &        4.4 &        4.3 &   5.03 &  $-0.73$  & $   5.5 \pm   1.1 $ \\
  101748322 & 2M19305707-0008228 &   $ 5.55 \pm  0.03 $ &   $  53.17 \pm  0.84  $ & OK   &    5.40 &     52.4  &      2 &       3 &   4710 & $-0.14 \pm 0.03 $ & $ 0.17 \pm 0.01 $ &    $ 1.34 \pm  0.08 $ &        5.5 &        4.5 &   6.93 &  $-0.19$  &                     \\
  102673776 & 2M06430619-0103534 &   $ 2.23 \pm  0.05 $ &   $  16.83 \pm  0.77  $ & OK   &    2.23 &     16.8  &      0 &       4 &   5070 & $-0.61 \pm 0.04 $ & $ 0.29 \pm 0.02 $ &    $ 1.69 \pm  0.28 $ &        0.8 &        0.8 &  14.05 &  $-0.25$  & $   4.3 \pm   4.2 $ \\
  102733615 & 2M06442450-0100460 &   $ 3.33 \pm  0.09 $ &   $  30.93 \pm  1.85  $ & poor &    3.06 &     30.9  &      1 &       4 &   4760 & $-0.29 \pm 0.04 $ & $ 0.16 \pm 0.02 $ &    $ 2.28 \pm  0.56 $ &        3.2 &        2.7 &  12.16 &  $-0.14$  & $   7.4 \pm   2.5 $ \\  
                                                                                                                                                                                                                                                         
  \hline                                                                                                                                                                                                                                                 
\multicolumn{18}{c}{\it \small $1\sigma$-outliers} \\
  \hline																															  
  100667041 & 2M19235081+0111425 &   $ 2.5  \pm  0.03 $ &   $  19.85 \pm  0.46  $ & OK   &    2.59 &     19.8  &      2 &       7 &   4400 & $-0.34 \pm 0.04 $ & $ 0.22 \pm 0.02 $ &    $ 1.23 \pm  0.11 $ &        4.1 &        6.3 &   5.13 &  $-0.53$  &                     \\
  100889852 & 2M19250803+0152285 &   $ 5.47 \pm  0.1  $ &   $  53.41 \pm  2.25  $ & poor &    5.62 &     56.1  &      2 &       7 &   4620 & $-0.32 \pm 0.04 $ & $ 0.27 \pm 0.02 $ &    $ 1.36 \pm  0.19 $ &        6.0 &        5.4 &   5.92 &  $-0.33$  & $   2.0 \pm   0.7 $ \\
  101029567 & 2M19255543+0014035 &   $ 2.33 \pm  0.04 $ &   $  17.76 \pm  0.81  $ & OK   &    2.35 &     17.7  &      0 &       7 &   4490 & $-0.65 \pm 0.04 $ & $ 0.28 \pm 0.02 $ &    $ 1.34 \pm  0.21 $ &        5.8 &        6.4 &   4.9  &  $-0.71$  & $   4.0 \pm   1.2 $ \\
  101073282 & 2M19261630+0116446 &   $ 5.63 \pm  0.09 $ &   $  56.33 \pm  2.01  $ & OK   &    5.53 &     56.9  &      0 &       3 &   4900 & $-0.08 \pm 0.03 $ & $ 0.10 \pm 0.01 $ &    $ 1.65 \pm  0.21 $ &        1.8 &        1.9 &   6.48 &  $-0.24$  & $   4.3 \pm   0.5 $ \\
  101200652 & 2M19270430+0120124 &   $ 2.23 \pm  0.04 $ &   $  17.11 \pm  0.70  $ & poor &    2.36 &     17.5  &      1 &       7 &   4500 & $-0.59 \pm 0.04 $ & $ 0.19 \pm 0.02 $ &    $ 1.38 \pm  0.23 $ &        3.8 &        6.2 &   5.27 &  $-0.55$  & $   3.7 \pm   1.0 $ \\
  101364068 & 2M19281113-0020004 &   $ 2.83 \pm  0.04 $ &   $  21.34 \pm  0.68  $ & OK   &    2.84 &     22.2  &      0 &       3 &   4650 & $-0.21 \pm 0.04 $ & $ 0.21 \pm 0.01 $ &    $ 1.30 \pm  0.15 $ &        8.0 &        5.9 &   6.03 &  $-0.38$  & $   2.8 \pm   0.5 $ \\
  101392012 & 2M19282435+0117076 &   $ 1.47 \pm  0.02 $ &   $   9.12 \pm  0.25  $ & OK   &    1.50 &      9.06 &      1 &       3 &   4390 & $-0.65 \pm 0.04 $ & $ 0.26 \pm 0.02 $ &    $ 1.10 \pm  0.10 $ &        6.2 &        7.6 &   5.35 &  $-0.55$  & $   4.5 \pm   1.0 $ \\
  101419125 & 2M19283555-0013131 &   $ 6.40 \pm  0.09 $ &   $  63.78 \pm  2.14  $ & poor &    6.57 &     65.8  &      1 &       7 &   4810 & $-0.49 \pm 0.04 $ & $ 0.27 \pm 0.02 $ &    $ 1.25 \pm  0.15 $ &        5.9 &        5.8 &   5.58 &  $-0.50$  & $   4.9 \pm   0.8 $ \\
  101476920 & 2M19285918+0036543 &   $ 2.20 \pm  0.03 $ &   $  16.63 \pm  0.54  $ & OK   &    2.25 &     17.3  &      0 &       7 &   4410 & $-0.14 \pm 0.03 $ & $ 0.14 \pm 0.01 $ &    $ 1.45 \pm  0.16 $ &        5.1 &        4.8 &   5.17 &  $-0.64$  &                     \\
  101665008 & 2M19302198+0018463 &   $ 6.02 \pm  0.05 $ &   $  61.93 \pm  1.35  $ & OK   &    6.30 &     62.6  &      1 &       3 &   4600 & $-0.06 \pm 0.03 $ & $ 0.16 \pm 0.01 $ &    $ 1.28 \pm  0.15 $ &        4.3 &        6.4 &   6.88 &  $-0.20$  & $   7.1 \pm   0.3 $ \\
  102768182 & 2M06451106-0032468 &   $ 2.94 \pm  0.06 $ &   $  27.18 \pm  1.29  $ & poor &    2.94 &     27.0  &      0 &       3 &   4840 & $-0.29 \pm 0.04 $ & $ 0.11 \pm 0.02 $ &    $ 2.17 \pm  0.37 $ &        1.7 &        1.7 &  10.25 &  $-0.05$  & $  10.3 \pm   0.5 $ \\
  
  \hline
\hline
\end{tabular}
}
\tiny{\tablefoot{\tiny{
$^a$ Quality of CoRoT lightcurve and the automated global fits; 
$^b$ Results of individual supervised fit to the lightcurves; 
$^c$ Flag on supervised fits ($0 =$ automated and supervised fit are consistent within 1$\sigma$. $1=$ there are two possible solutions
for $\Delta\nu$ or $\nu_{\mathrm{max}}$, due to the ambiguity of radial and dipole oscillation modes. $2=$ supervised fit yields improved results); 
$^d$ Overall uncertainties: $\sigma{T_{\mathrm{eff}}}=91$ K \citep{Holtzman2015}; 
$^e$ 1$\sigma$ age upper limit, using the seismic results from the automatic pipeline; 
$^f$ 1$\sigma$ age upper limit, using the seismic results from the supervised seismic analysis; 
$^g$ Typical uncertainties: $\sim 0.1$ kpc, for the most distant stars in LRa01 $\sim0.5$ kpc; 
$^h$ Typical uncertainties: $<0.1$ kpc, for the most distant stars in LRc01 $\sim0.4$ kpc; 
}}}
\end{sidewaystable*}
\end{appendix}
\end{document}